# InN nanowire solar cells on Si with amorphous Si interlayer deposited by sputtering

M. Sun[1,*], R. Gómez[1], B. Damilano[2], J. M. Asensi[3,4], F. B. Naranjo[1] and S. Valdueza-Felip[1]

*[1]Photonics Engineering Group. University of Alcala, 28871 Alcalá de Henares, Spain.*
*[2]Université Côte d'Azur, CNRS, CRHEA, Valbonne, France.*
*[3]Department de Física Aplicada, Universitat de Barcelona, 08028, Barcelona, Spain.*
*[4]Institut of Nanoscience and Nanotechnology (IN2UB), Universitat de Barcelona, 08028, Barcelona, Spain.*
*michael.sun@uah.es

## Abstract

Here, we report the first experimental demonstration of InN nanowire solar cells deposited by RF sputtering with a bandgap energy of 1.78 eV. By adding an amorphous Si (a-Si) buffer to the *n*-InN/*p*-Si structure, we have improved the photovoltaic performance of the resulting devices while maintaining their material quality. We have firstly optimized the deposition of Si on Si(100) by DC sputtering, obtaining an amorphous material with bandgap energy of 1.39 eV. Then we have studied the influence of the thickness of the a-Si buffer layer (0-25 nm) on the structural, morphological, electrical, and optical properties of InN nanowires on Si(100) substrates. With the use of a 15-nm buffer, *n*-InN/a-Si/*p*-Si nanowire heterojunction solar cells exhibit a promising short-circuit current density of 17 mA/cm$^2$, open circuit voltage of 0.37 V and fill factor of 35.5%, pointing to a power-conversion efficiency of 2.3% under 1-sun (AM 1.5G) illumination. These work demonstrates that the combination of in-situ sputtered a-Si, which could serve as potential passivation layer, and the light trapping enhancement by the nanostructured active layer leads to an improvement of the photovoltaic efficiency of sputtered III-nitride devices.

# 1. Introduction

Photovoltaic technology is crucial in the transition towards a sustainable energy future. As the world's population and energy demands continue to grow, it is becoming increasingly important to find alternative energy sources that are clean, efficient, and renewable. Solar energy is one such source that has the potential to meet a significant portion of our energy needs. III-nitrides have a wide range of material properties that are excellent for photovoltaic research. They show high electron mobility and can efficiently convert solar energy into electricity thanks to their direct bandgap that covers the UV-NIR range from 6.2 eV for AlN to 0.7 eV for InN. Additionally, III-nitrides have excellent thermal and chemical stability, making them durable and long-lasting for use in harsh environments. Furthermore, III-nitrides have already found successful commercial applications in the production of LEDs and lasers, demonstrating their potential for use in other areas of optoelectronics. As such, exploring the potential of III-nitride materials for photovoltaic applications could lead to significant advancements in solar energy technology.

During last decades, because of the exponential development and maturation of the technology, many III-nitride based devices have been introduced to the market, such as high electron mobility transistors (HEMTs), light-emitting diodes [1] (LEDs) and laser diodes (LDs) [2]. InN has attracted much attention in the last few years due to its several important attributes, including high absorption coefficient, high carrier mobility and large drift velocity that are required for high efficiency photovoltaics [3-7]. Several reports of InN layers deposited by MOCVD [8-10], MBE [11-12] and reactive sputtering [13-22] have been published despite the difficulties in obtaining good quality InN. Substantial effort has been dedicated to exploring III-nitride nanowires [23-26] and nanocolumns [27-31] due to their distinct structural and morphological characteristics, such as high aspect ratios, large surface-to-volume ratios, and the ability to tune their dimensions. These features enable enhanced light absorption, improved carrier collection and reduced defects, which can raise the overall performance of InN solar cells. Previous research groups have demonstrated the deposition of InN layers via radio-frequency (RF) sputtering on different types of substrates, mainly on sapphire and Si(100) and have optimized the deposition conditions to obtain high material quality and good electrical performance on InN-Si(100) heterojunctions devices [32-36]. With the aim of improving the photovoltaic performance of these devices, we will study the influence of introducing an a-Si layer on the InN/Si interface acting both as a potential surface passivation layer for silicon [37], and buffer layer.

## 2. Experimental details

Silicon and InN layers were deposited using a reactive (DC and RF, respectively) magnetron sputtering system (AJA International, ATC ORION-3-HV) on *p*-doped 375-µm-thick Si(100) with resistivity of 1-10 Ω·cm and on sapphire substrates. This system is equipped with 2-inch confocal magnetron cathodes of *p*-type B-doped pure Si (99.999 %) and In (99.995 %) targets. The background pressure of the system was in the order of $10^{-5}$ Pa and the distance between target and substrate was fixed at 10.5 cm. Substrates were chemically cleaned in organic solvents and outgassed inside the chamber for 30 minutes at 550 ºC. Before deposition, the surface of the targets and substrates were cleaned using a soft plasma etching with Ar (2 sccm and 20 W).

Firstly, we study the deposition of silicon on Si(100) substrates. Silicon thin films were deposited in a pure Ar (99.9999%) atmosphere (Ar flow of 2 sccm) and a pressure of 0.47 Pa during 2 hours at 550 ºC with different DC powers, namely 30, 40, 50 and 60 W. These deposition parameters lead to compact Si films with an amorphous phase [38].

Then, a set of InN samples was grown using the previously reported a-Si thin film [38] as buffer layer at 30 W of DC power. The nominal thickness of the buffer was varied as 0 (no buffer), 4, 8, 15 and 25 nm (samples B0, B4, B8, B15 and B25, respectively, where the number stands for the thickness of the buffer layer). The InN layers were deposited in pure $N_2$ (99.9999%) atmosphere ($N_2$ flow of 14 sccm) and a pressure of 0.47 Pa. The RF power applied to the In target, growth temperature and deposition time were fixed at 30 W, 550 ºC and 150 min, respectively.

The crystalline orientation and mosaicity of the films were evaluated by high-resolution X-ray diffraction (HRXRD) measurements using a PANalytical X'Pert Pro MRD system. The thickness and morphology of the layers were studied by field-emission scanning electron microscopy (FESEM) and atomic force microscopy (AFM), respectively. Meanwhile, Hall-Effect and optical transmission measurements were carried out at room temperature on samples co-deposited on sapphire to study their electrical and optical properties of the films.

Photothermal deflection spectroscopy (PDS) was utilized to investigate the semiconductor quality of the a-Si film utilized as a buffer layer. PDS is a precise technique capable of measuring absorption in the weak absorption region (photon energies hν below the bandgap $E_g$), which normally cannot be accessed by conventional optical measurements. In the utilized PDS system, the surface of the a-Si sample that was deposited onto sapphire was exposed to monochromatic light ranging from 400-2000 nm. This served as the pumping source at a perpendicular angle, with a chopping frequency of 4 Hz. As a probe, a 660 nm semiconductor laser was used to scan the sample surface. The position of the laser probe was then deflected based on the thermal energy conversion resulting from nonradiative recombination of excited electrons induced by the pumping light. Details of the PDS system can be found in [39].

Devices with ~1 $cm^2$ area were processed using the developed structures with the *n*- and *p*-type contacts formed by 120 nm-thick Al deposited by DC sputtering. The *p*-type contact was annealed at 450 ºC for 3 minutes under nitrogen atmosphere to form an ohmic contact [32]. They were characterized by current density-voltage (J-V) curves using a 2-point probe station coupled with a source meter unit in dark and under 1 sun (1 $kW/m^2$) of standard illumination (solar simulator class AAA with AM1.5G spectrum) [40]. Their spectral response was measured using a 250 W halogen lamp coupled with a monochromator and calibrated with a reference Si photodetector. A 518-nm laser with an output power of 0.8 mW was used to precisely determine the responsivity (mA/W) and the external quantum efficiency (EQE) of the cells at that wavelength.

# 3. Results and discussion

## 3.1 Deposition of silicon thin films on Si(100)

Our group has reported the development of amorphous Si (a-Si) on sapphire and GaN-template substrates by DC sputtering at substrate temperature from RT to 550 ºC [38]. In ref. 38, we report that the a-Si deposited on sapphire substrate shows a smooth surface (rms of below 1 nm) and it was later measured by resistivity measurements obtaining high resistivity films (~165-250 Ω.cm).

In this study, we investigate the thickness, morphology, and surface roughness of the deposited amorphous Si on Si(100) substrates. Figure 1 shows the SEM and AFM images of the a-Si films deposited at different DC power (30, 40 and 50W). In all cases, a compact and smooth Si film is grown regardless the power applied to the target. Additionally, the root-mean-square (rms) surface roughness remains below 1 nm for all the analysed samples. Finally, the deposition rate increases from ~70 nm/h for 30 W of DC power to ~85 nm/h for 40 W and finishing with ~120 nm/h for 50 W of DC power. Moreover, these Si films show a high resistivity. All these layer properties agree with the obtained ones for a-Si deposited on sapphire sample [38]. From here on, we will use 30 W of DC power for the growth of Si to guarantee the slowest growth rate allowing a better control of the thickness for its use as buffer layer.

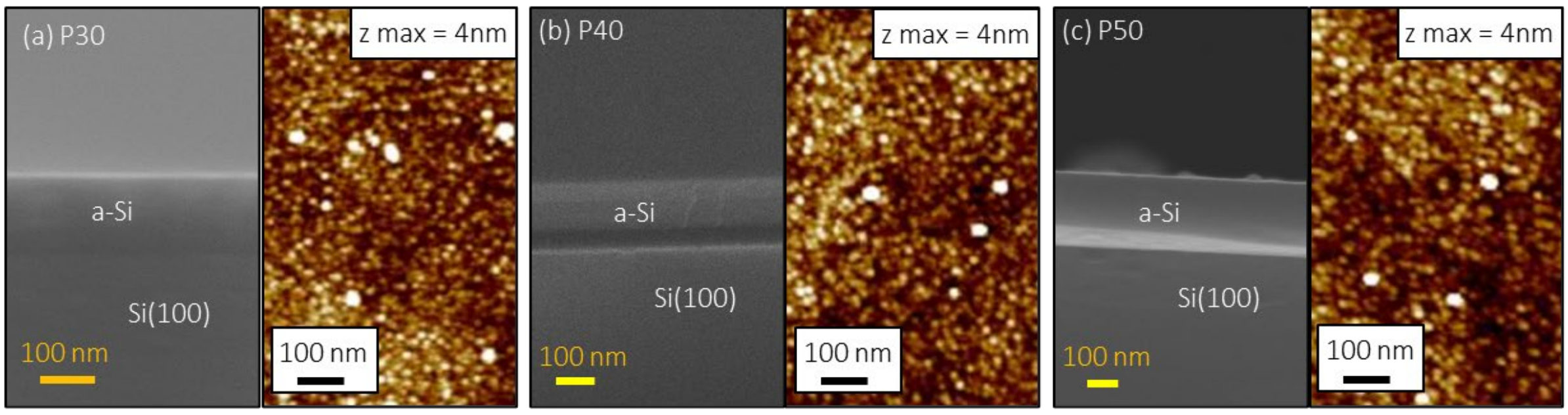

**Figure 1.** SEM and AFM images of a-Si deposited on Si(100) substrates at different DC powers.

Figure 2 shows the absorption coefficient spectra α(hν) of an a-Si sample deposited on a sapphire substrate at 30 W DC. The α value was determined by fitting the experimental PDS absorption data with the theoretical absorption of the substrate/layer structure, as deduced through the transfer matrix method (TMM) [39]. The exact film thickness (148 nm) and the spectral dependence of the refractive index were obtained by analyzing conventional optical transmittance and reflectance measurements of the same PDS sample. Optical measurements are also valuable for calibrating the PDS signal as the PDS absorbance needs to align with the optical absorbance in the spectral region of strong absorption. Finally, fitting the absorption coefficient spectrum α(hν) to that predicted by the Urbach-Tauc model allowed us to determine the bandgap value $E_g$ and the Urbach energy $E_U$ of the a-Si sample [41].

The analytical expression of Tauc-Urbach model is:

$$\alpha(h\nu) \propto \begin{cases} \dfrac{\left(h\nu - E_g\right)^2}{h\nu} & \text{if } h\nu > E_1 \\ \exp\dfrac{h\nu}{E_U} & \text{if } h\nu \leq E_1 \end{cases}$$

where $E_1$ is a transition energy (greater than $E_g$) that separates the Tauc behavior from the Urbach behavior. Note that by constraining α(hν) in $E_1$ to vary continuously and smoothly, the model has only 3 independent parameters ($E_g$, $E_U$ and $E_1$).

By fitting the experimental data with the Urbach-Tauc model, a value of 1.39 eV is obtained for the energy of the bandgap $E_g$. A very similar value of 1.4 eV was reported by Hossain [42] in similar non-hydrogenated a-Si layers. In addition, the transition energy $E_1$, which determines the onset of exponential Urbach absorption, was found to be 1.64 eV and the Urbach energy Eu was found to be 138 meV. The relatively high value of Eu suggests considerable structural disorder in the a-Si film.

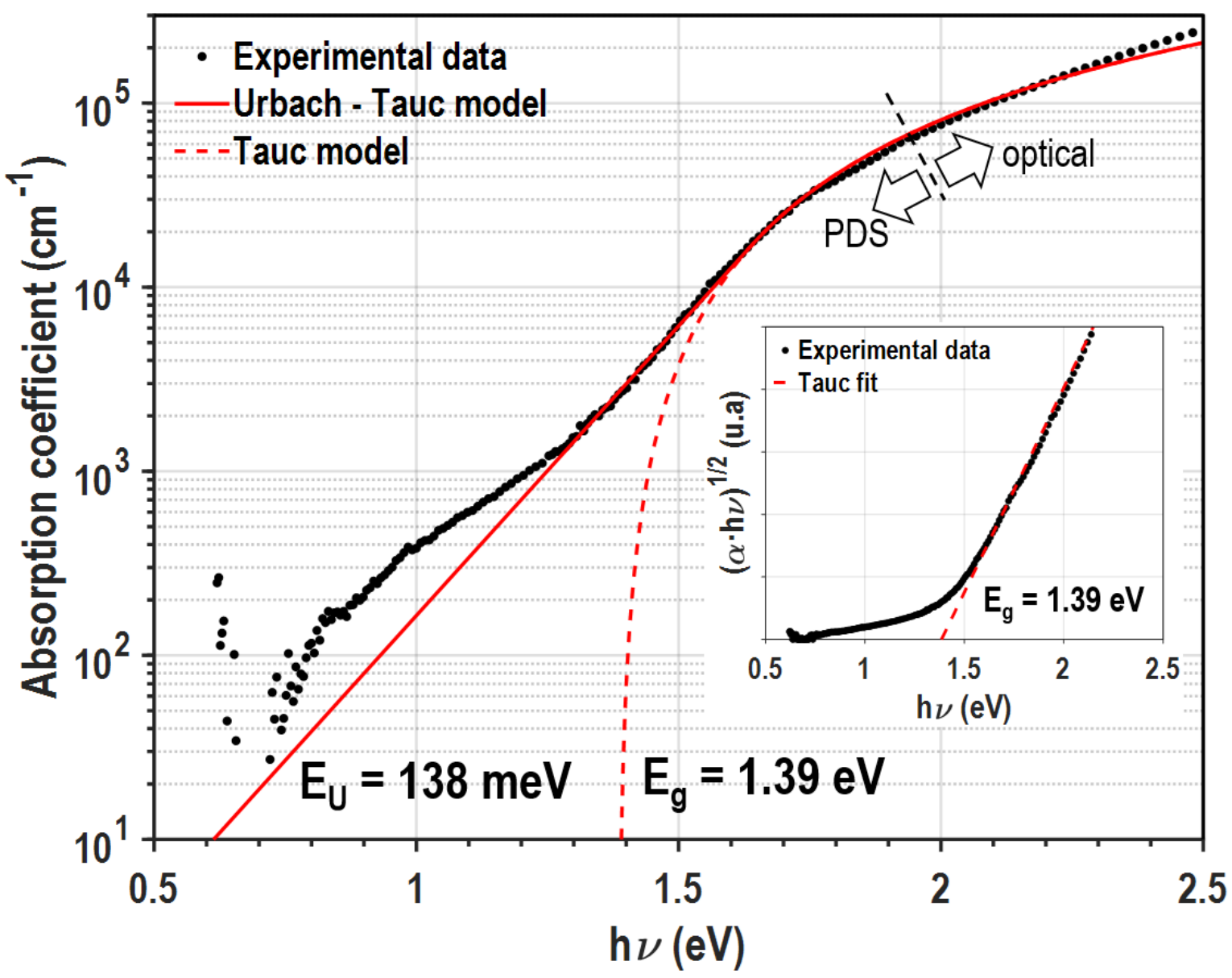

**Figure 2.** Absorption coefficient as a function of the photon energy hν of an a-Si layer deposited on sapphire at 30 W DC. The absorption coefficient is obtained from the PDS measurement for hν below 1.95 eV. The solid red line represents the fit using the Urbach-Tauc model and the dashed red line represents the portion of the fit corresponding to the simple Tauc model (the inset displays the standard Tauc plot).

## 3.2 InN thin films on amorphous Si interlayer

### 3.2.1 Structural and morphological characterization

A set of samples of InN layers deposited on Si(100) and sapphire was grown to study the effect of the a-Si buffer layer, using different thickness of buffer (samples B0, B4, B8, B15 and B25, were the number stands for the thickness of the buffer layer). Figure 3 show the HRXRD 2θ/ω scans carried out on samples grown on Si(100). X-ray diffraction data of the InN/a-Si/Si(100) structures show a wurtzite structure oriented along the *c*-axis for all samples. The FWHM of the 2θ/ω diffraction pattern of the (0002) InN reflection remains in the range of 0.23° to 0.3° with the introduction of the a-Si buffer, pointing to InN material with a similar structural quality with and without buffer layer. On the other hand, the FWHM of the rocking curve of the (0002) InN diffraction peak (not shown), which is linked to its mosaicity, remains in the range of 6.8° to 8.2° as shown in Table 1. The exception is the sample with 25 nm buffer layer, showing a FWHM of 6.9°, in agreement with the reduction of this parameter for thick buffer layer observed by Monteagudo-Lerma *et al.* [43].

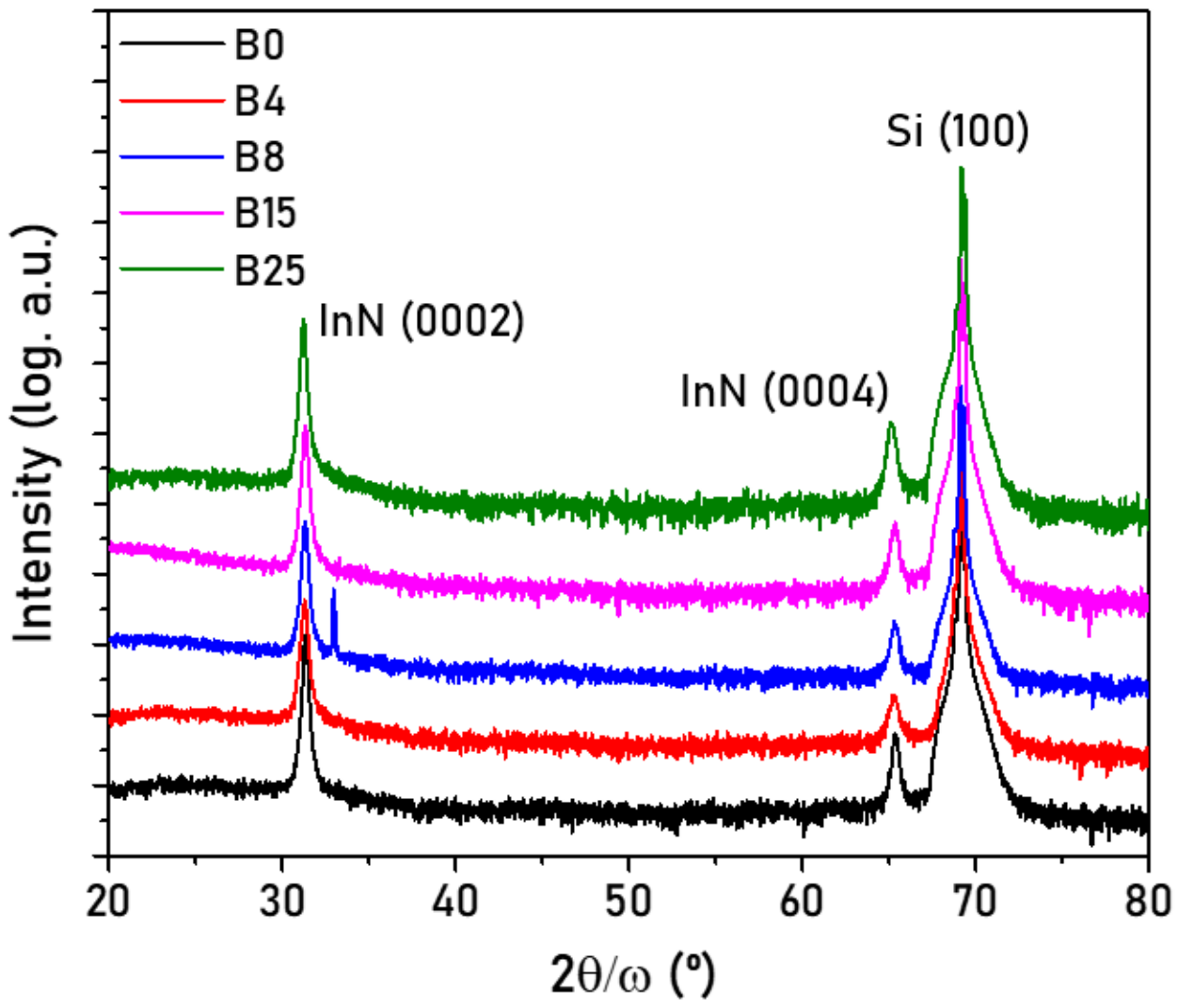

**Figure 3.** 2θ/ω XRD diffractograms of InN on Si(100) with different Si buffer thickness.

Figure 4 shows the SEM and AFM images of the InN samples deposited on Si(100) with different a-Si buffer. The samples show a columnar morphology (see Figure 4), as reported in previous works studying InN films deposited by RF sputtering [43]. AFM measurements confirm the columnar morphology of the samples, with a column density of 110 μm$^{-2}$ for B0 and increases to ≈165-190 μm$^{-2}$ when adding a buffer layer. At the same time, column diameter decreases from ≈86 nm for B0 to ≈60-69 nm with buffer layer. The thicknesses of the InN samples, extracted from SEM images, is summarized in Table 1.

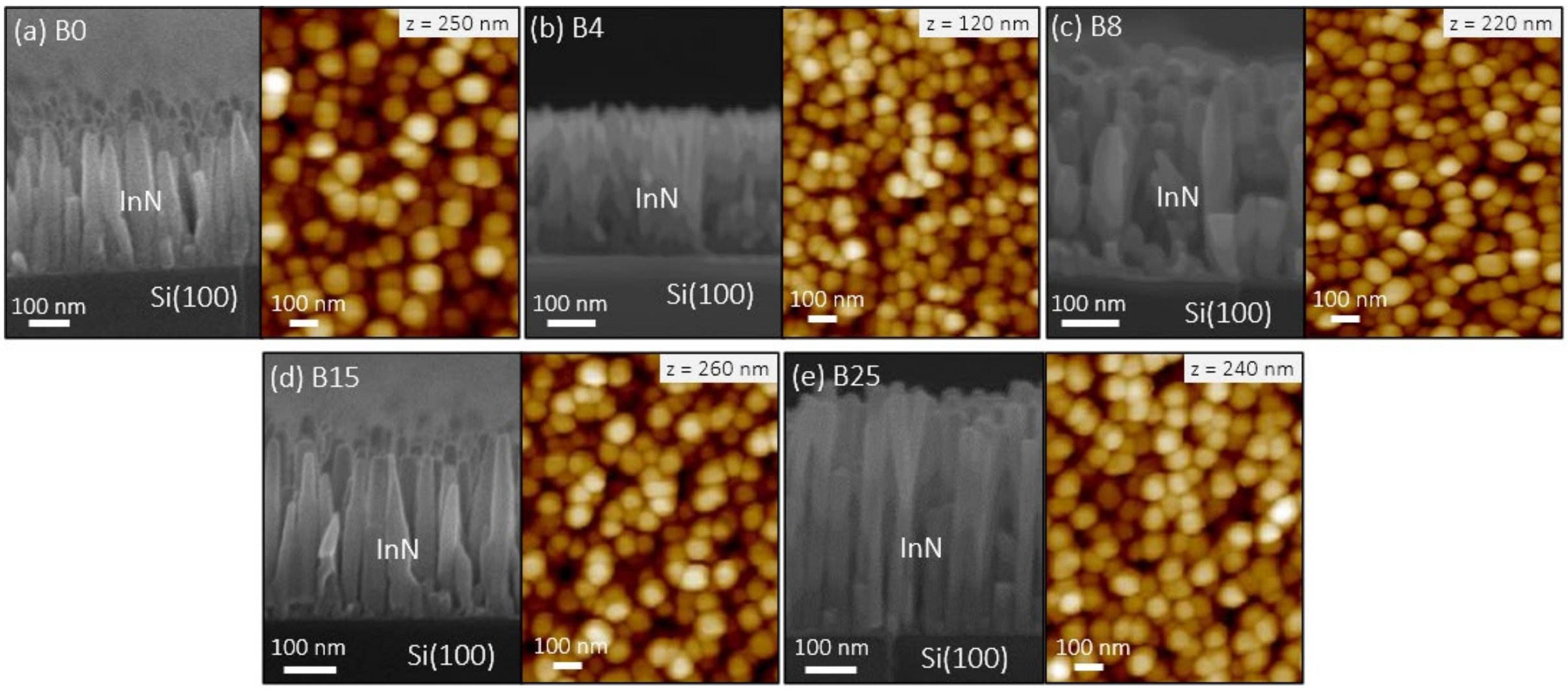

**Figure 4.** Cross-section SEM (left) and AFM (right) images of InN on Si(100) samples with (a) 0 nm (B0) (b) 4 nm (B4) (c) 8 nm (B8) (d) 15 nm (B15) and (e) 25 nm (B25) of-Si buffer layer. Note: the "z" in the AFM images refers to the vertical axis.

**Table 1.** Summary of the morphological properties of InN samples on Si(100) with different a-Si buffer thickness.

| Sample | a-Si buffer thickness [nm] | InN nanowire length on Si(100)* [nm] | FWHM rocking curve (°) | Column density** [μm$^{-2}$] | Column diameter [nm] |
|---|---|---|---|---|---|
| **B0** | 0 | 305 | 7.9 | 110 | 86 ± 12 |
| **B4** | 4 | 305 | 7.9 | 190 | 60 ± 12 |
| **B8** | 8 | 325 | 7.5 | 190 | 69 ± 12 |
| **B15** | 15 | 265 | 8.2 | 160 | 66 ± 15 |
| **B25** | 25 | 485 | 6.9 | 165 | 63 ± 10 |

* std error ± 5 nm ** std error ± 5 μm$^{-2}$

The highest nanowire length, observed for the sample with a 25 nm a-Si buffer, could be attributed to a difference in nucleation process induced by the low thermal conductivity of a-Si compared to crystalline Si (around 100 times lower [44]), leading to an increase of nucleation sites density. This, in turn, might promote a vertical growth mode, resulting in longer nanowires. In this sense, the column diameter in sample B25 does not exhibit a reduction along its length compared to the other samples, which display a pyramidal-like diameter structure. This structural difference explains the similar column diameters observed at the top surface, as corroborated by the AFM images. Changes in surface properties due to a buffer layer were also observed previously on AlInN samples grown on Si(100) by Núñez-Cascajero *et al* [45].

### 3.2.2 Electrical characterization

Hall-Effect measurements were conducted to investigate the electrical characteristics of InN samples deposited on sapphire substrates, which results are summarized in Table 2. It is worth noting that InN on sapphire samples without buffer layer show a compact morphology. However, they become columnar when adding an a-Si buffer layer, showing a similar morphology to the one of InN/Si samples of Fig. 4. Figure 5 shows the SEM images of InN samples deposited on sapphire without (a) and with 15-nm of buffer layer (b). This phenomenon regarding the observation of nanowire structures on sapphire substrates, can be attributed to the effect of the buffer layer in the growth mechanism of the InN film, as observed in previous studies with AlN buffer, where the morphology of the InN samples changed depending on the buffer thickness [43].

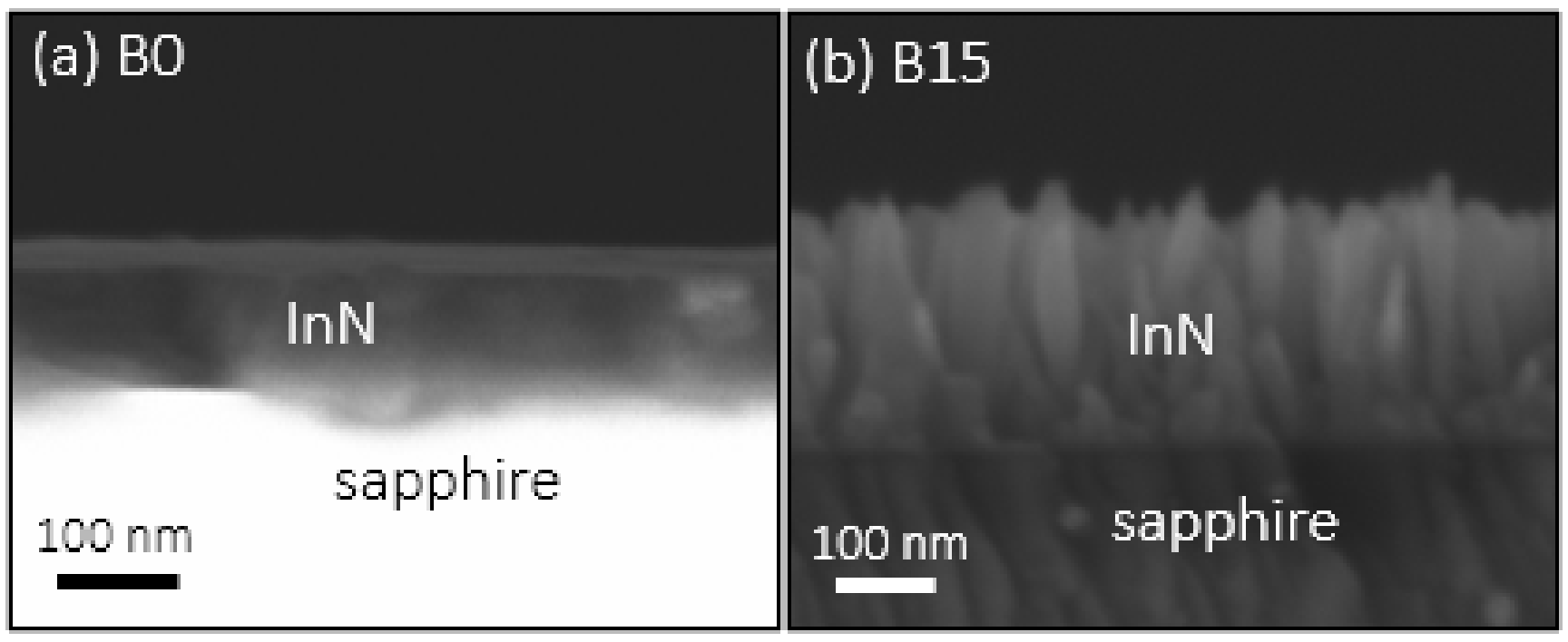

**Figure 5.** Cross-section SEM images of InN on sapphire samples with (a) 0 nm (B0) and (b) 15 nm (B15) of-Si buffer layer.

First, the resistivity of the layer increases almost a factor of 10 when adding 4 nm of a-Si buffer. This is most probably due to the change in morphology as stated before, changing from a compact layer to a nanocolumnar layer characterized by its vertically aligned structures. These structures possess a larger number of interfaces compared to a compact layer, as each column seems to be isolated from the rest (except for the region close to the substrate). Increased surface and interface scattering contribute to the observed higher resistivity. Nevertheless, a path conduction at the interface between nanowires and buffer/substrate can be considered.

However, as we increase the buffer thickness the resistivity slightly rises to 15 nm and finally it jumps five times for 25 nm of buffer. This behaviour is likely attributed to the high resistivity or insulating properties of the amorphous silicon (a-Si) buffer and the reduction of the parallel conduction path at the interface between nanowire and buffer layer due to the change in nucleation process observed in the sample. We should note that this nanocolumnar morphology can also affect the resistivity measurements of these samples.

**Table 2.** Summary of the electrical and optical properties of InN samples on sapphire with different a-Si buffer thickness.

| Sample | a-Si buffer thickness [nm] | InN nanowire length on sapphire* [nm] | Resistivity [mΩ.cm] | Bandgap energy [eV] |
|---|---|---|---|---|
| **B0** | 0 | 130 | 0.19 | 1.78 |
| **B4** | 4 | 265 | 1.73 | 1.85 |
| **B8** | 8 | 235 | 1.80 | 1.88 |
| **B15** | 15 | 240 | 2.12 | 1.93 |
| **B25** | 25 | 400 | 9.54 | 1.87 |

* std error ± 5 nm

On the other hand, carrier concentration and mobility values cannot be reliably obtained through Hall Effect measurements for samples with the a-Si buffer layer due to their columnar morphology. For the compact InN sample without buffer layer, Hall-Effect measurements reveal a carrier concentration of $1x10^{21}$ $cm^{-3}$ and a carrier mobility of 36 $cm^2/V.s$.

### 3.2.3 Optical characterization

The optical properties of the layers were first studied by transmittance measurements performed on samples deposited on sapphire. Figure 6 shows the transmittance spectra of the samples under study. The absorption of the layers can be derived from transmission spectra following the relation $\alpha(E) \propto -ln(T)$, without considering optical scattering and reflection losses. The absorption coefficient can be fitted with the sigmoidal equation $\alpha = \frac{\alpha_0}{1+e^{\frac{E_0-E}{\Delta E}}}$ , where $\alpha$ is the linear absorption, $E$ is the photon energy, $E_0$ is the mean energy identified as the "effective bandgap" of the material and $\Delta E$ is the absorption band edge broadening. The calculated sigmoidal approximation of the absorption coefficient is used to estimate the bandgap energy from the linear fit of the squared absorption coefficient ($\alpha^2$), as plotted in the inset of Figure 6. The samples show an apparent bandgap energy of $\approx$1.8 eV (696 nm) for the InN without buffer layer and then a blue-shift of the bandgap energy to $\approx$1.9 $\pm$ 0.5 eV (652 nm) is observed when introducing the a-Si buffer layer (see Table 1). The observed increased in optical bandgap could be tentatively attributed either to the formation of $In_2O_3$ resulting from the oxidation of the layer, as observed by Motlan et. al. [46], or by an increase in carrier concentration of the InN nanocolumns and the induced Burstein-Moss effect [47].

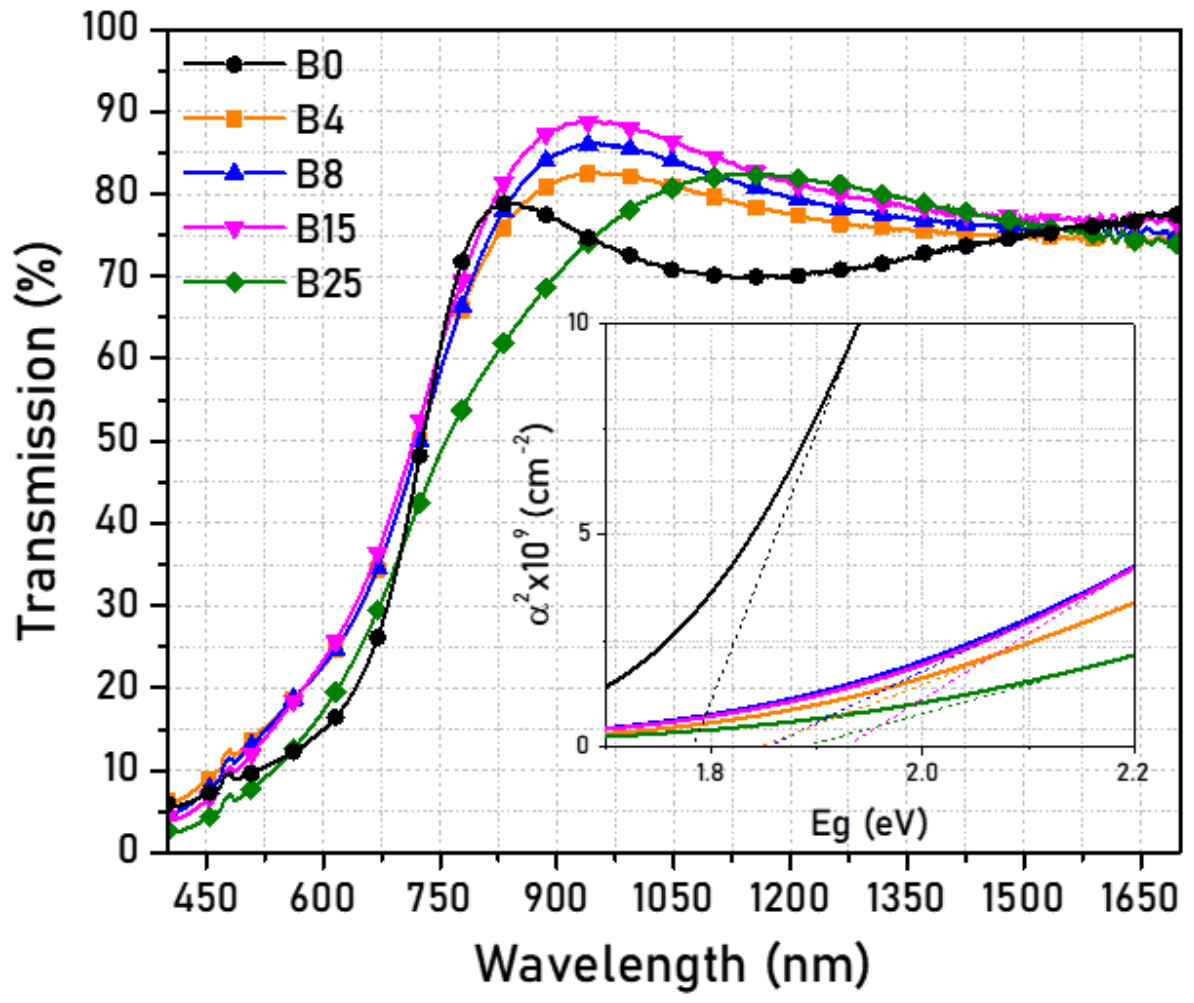

**Figure 6.** Transmission spectra of InN samples on sapphire substrates. Inset: squared absorption coefficient $\alpha^2$ vs the energy extracted from the sigmoidal approximation of the InN layers grown on sapphire. Dashed lines are the linear fits used to estimate the apparent optical band gap energy of the samples $E_g^{Abs}$.

### 3.3 Photovoltaic performance of InN on Si(100) devices with a-Si buffer

Figure 7 shows the J-V measurements of InN/a-Si/Si(100) devices as a function of the a-Si buffer thickness (a) in the dark and (b) under 1 sun AM 1.5G illumination. The inset of Fig. 7(a) shows a schematic structure of the devices under study. Dark J-V curves were performed to evaluate the behaviour of the device as a diode. Devices in the dark show a diode rectifying behavior with a ratio of ~500 measured at -1 and 1 V for the B4 device, which is attributed to the electric field of the *p-n* heterojunction. The curves were analysed using the one diode model [48] to extract the series and shunt resistances ($R_S$, $R_{SH}$), the reverse saturation current density ($J_0$) and the ideality factor ($\eta$). Measured and fitted curves are shown in Figure 7(a). Moreover, a clear photo-response has been consistently observed. J-V curves measured under illumination (shown in Figure 6(b)) served to evaluate the photovoltaic characteristics of the devices in terms of open circuit voltage ($V_{OC}$), short-circuit current density ($J_{SC}$) and Fill Factor (FF). The power conversion efficiency was obtained as $Eff.(\%) = \frac{VocJsc\,FF}{P_{in}} \times 100$, being $P_{in}$ = 1 kW/m$^2$ for the incident sunlight. All results of the analysis are summarized in Table 3.

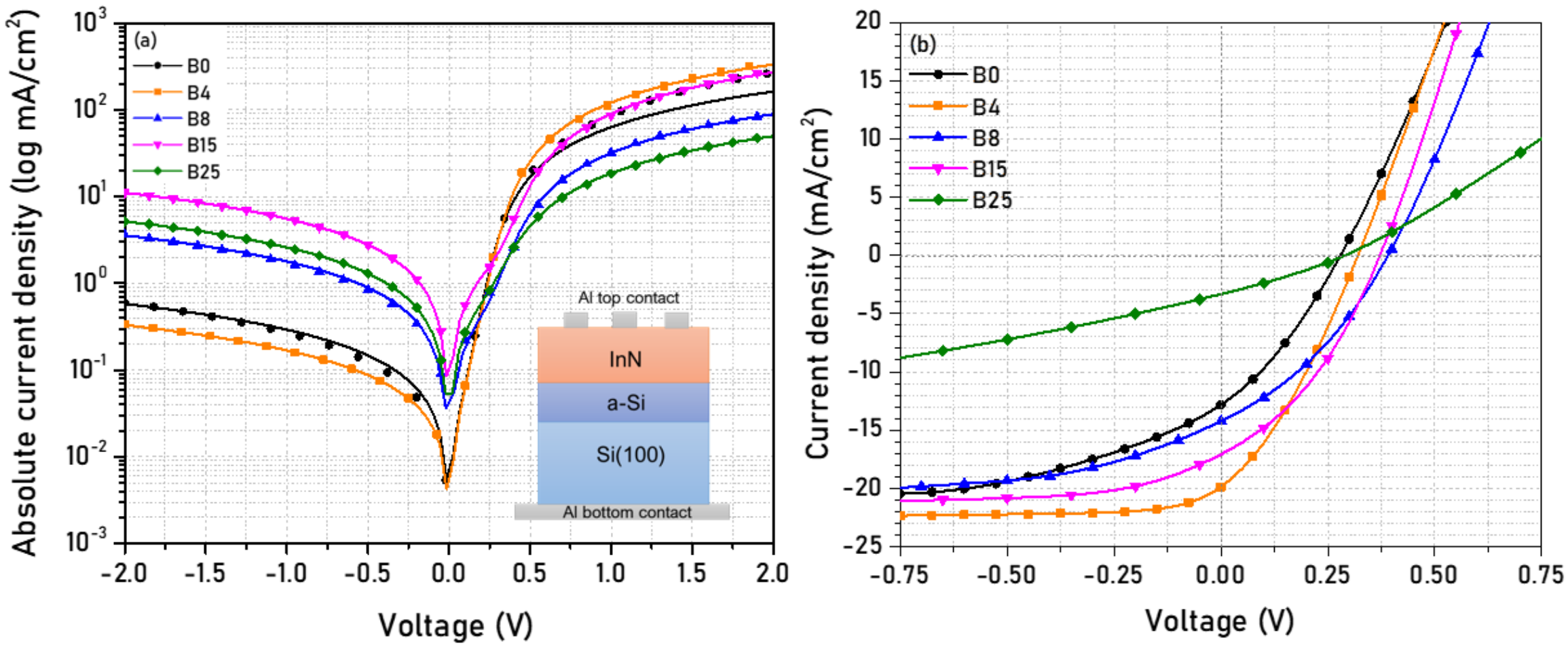

**Figure 7.** J-V measurements of InN/a-Si/Si(100) devices as a function of the a-Si buffer thickness (a) in the dark, where symbols correspond to experimental data and lines to the fit extracted from the one-diode model [48], and (b) under 1 sun AM 1.5G illumination. Inset of (a) is a schematic of the devices under study.

The series resistance in a solar cell can be mainly due to the internal resistance of the device against the movement of carriers through the structure, and the top and bottom contact resistances; in our case we may have a mix of both causes. The main impact of an increase of $R_S$ is to reduce the short-circuit current of the cell. In this study, the series resistance decreases when incorporating an a-Si buffer layer of 4 to 15 nm thickness, leading to higher $J_{SC}$, compared to a device without buffer (B0). However, for 25 nm of buffer thickness, $R_S$ increases leading to a drop in $J_{SC}$, and therefore in the power conversion efficiency. This could be caused due to the resistive buffer, which would be thick and resistive enough to difficult the extraction of the charges from the depletion region [49].

Undesired low shunt resistance usually causes power losses in solar cells by providing an alternate path for the light-generated current, reducing thus the open-circuit voltage of the device. In this study, this trend with buffer layer thickness is not so clear. The shunt resistance increases when we incorporate an a-Si buffer layer of 4 nm, leading to higher $V_{OC}$, compared to a device without buffer (B0). At the same time, the strong drop that suffers $R_{SH}$ when we increase the buffer thickness above 4 nm is not decreasing the $V_{OC}$. On the contrary, it increases for 8-15 nm thickness devices, and it reduces for 25-nm ones. This result is tentatively attributed to a stronger effect of the series resistance than to the shunt resistance one, on the drop of the performance of the B25 device. In our study, the observed $V_{OC}$ values of 0.28 - 0.40 eV agrees with the ones obtained from simulations [50] of InN on Si heterojunctions with a bandgap of 1.9 eV and a carrier concentration in the order of $10^{21}$ $cm^{-3}$.

The reverse saturation current and ideality factor present values in the same order of magnitude for all the analysed devices, showing no clear difference between devices with and without a-Si buffer, neither a clear dependency with the buffer thickness.

**Table 3.** Summary of the electrical and photovoltaic characteristics of the InN/a-Si/Si(100) devices. The area was calculated considering the shadow of the top contact (~0.13 $cm^2$).

| **Sample** | **Area ($cm^2$)** | **$R_S$ ($\Omega\cdot cm^2$)** | **$R_{SH}$ ($k\Omega\cdot cm^2$)** | **$J_0$ ($\mu A/cm^2$)** | **η** | **$V_{OC}$ (V)** | **$J_{SC}$ ($mA/cm^2$)** | **FF (%)** | **Eff. (%)** |
|---|---|---|---|---|---|---|---|---|---|
| **B0** | 0.84 | 9.7 | 3.5 | 3.7 | 1.6 | 0.28 | 12.8 | 31.4 | 1.1 |
| **B4** | 0.99 | 4.6 | 8.5 | 6.2 | 1.8 | 0.32 | 19.9 | 32.1 | 2.0 |
| **B8** | 0.75 | 16.5 | 0.5 | 1.1 | 1.8 | 0.39 | 14.2 | 33.7 | 1.9 |
| **B15** | 0.67 | 5.2 | 0.2 | 1.2 | 1.8 | 0.37 | 17.2 | 35.5 | 2.3 |
| **B25** | 0.75 | 30.4 | 0.4 | 1.9 | 1.9 | 0.29 | 3.3 | 28.8 | 0.3 |

The introduction of the a-Si buffer produces an increase of the $V_{OC}$, $J_{SC}$, FF and therefore of the efficiency of the analysed devices with 4-15 nm thickness. In particular, a two-fold improvement in efficiency from 1.1% to 2.3% is achieved when adding 15 nm of a-Si buffer layer to the *n*-type InN / *p*-type Si structure. It may be noted that these results surpass the ones previously obtained on *p-i-n* InN nanowires deposited by MBE on *n*-type Si(111), that achieved a maximum efficiency of 0.68% with a CdS passivation treatment [26]. In addition, other studies on InGaN nanowires can be found in the literature. A coaxial *n*-GaN/i-$In_xGa_{1-x}N$/*p*-GaN nanowire photovoltaic device was reported by Dong [51] *et al.* achieving efficiencies of 0.19%. Wierer [52] *et al.* investigated a large area radial III-nitride nanowire solar cell reporting efficiencies of 0.3%. Hybrid solar cells containing III-nitride nanowire and Si were also reported by Tang [53] *et. al.*, with *p*-GaN nanowire grown on *n*-Si substrate achieving 2.73% conversion efficiency. It should be pointed out that in this latter case, the nanowires are synthesised by chemical vapor deposition, using gold nanoparticles catalyst.

To better understand the functioning of the *p-n* heterojunction and determine the mechanism by which the introduction of the a-Si buffer layer affects the carrier transport through the structure of the device, we conducted simulations of the energy-band diagram using nextnano$^3$ software [54]. Higher-energy photons will be absorbed at the InN, whereas lower-energy photons will reach silicon and be then absorbed. The results reveal an increase of the separation between the notch formed in the conduction band at Si-III nitride interface and the valence band of the Si substrate (see inset of Figure 8(a)) when adding the a-Si interlayer. This separation plays a crucial role in mitigating the recombination rate though tunnelling effect at the interface. This phenomenon can be the responsible of the increase in the overall efficiency of the devices. At the same time, when increasing the a-Si thickness to 25 nm, a mitigation of the electric field inside the layer takes place, thus reducing the speed of the carriers and increasing the chances to recombine. This enhances the resistive effect of the amorphous Si layer resulting in an increase of $R_S$ and a drop in $J_{SC}$ and in the efficiency (see Table 3).

Finally, Figure 8(b) shows EQE measurements performed on devices without (B0) and with a 4-nm buffer (B4). They both own a very similar spectral shape with a maximum at ≈830 nm, and the cut-off related to Si at ≈1050 nm, respectively. The shape and EQE values of the devices correlate with results previously obtained in similar nanowire InN devices deposited by sputtering [32]. The drop of the EQE at short wavelengths occurs due to the high surface recombination rate and low minority carrier diffusion length of the InN layer. Peak EQE values at 830 nm increase from 63% for B0 and 76% for B4 devices, in accordance with J-V results. Moreover, a slight blue-shift of the cut-off related to the InN is observed when introducing the a-Si buffer layer, as displayed in the inset of Figure 8(b). We deduce that this small change could be due to the increase of bandgap energy from ≈1.8 (688 nm) to ≈1.9 eV (652 nm), as already discussed in the optical characterization section.

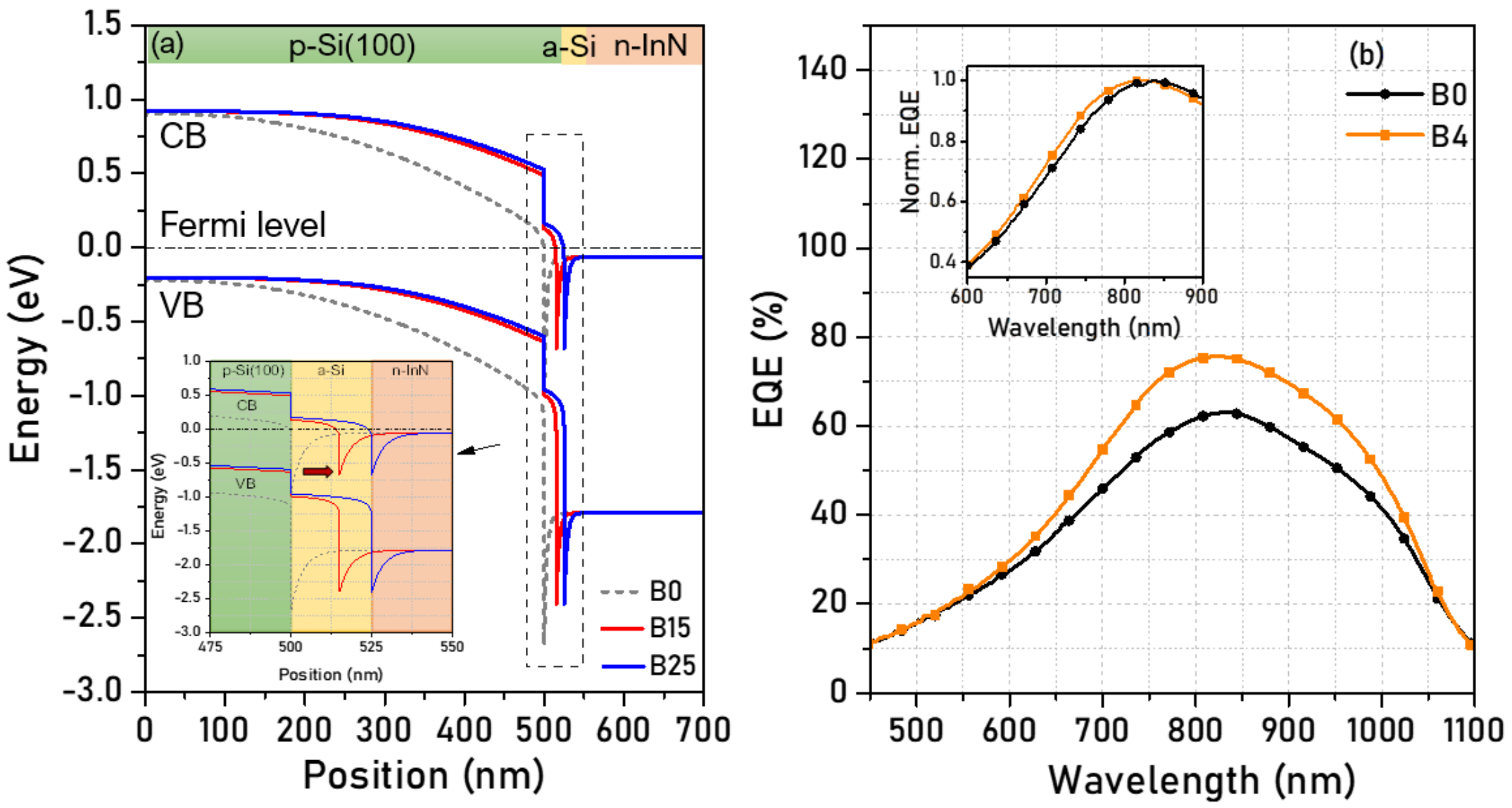

**Figure 8.** Influence of the a-Si buffer thickness on (a) electronic band structure of the device with 0, 15 and 25 nm of a-Si buffer and on (b) the responsivity of the InN on Si(100) devices.

A clear further amelioration of the devices is needed in order to increase their efficiency. In this sense, the open-circuit voltage could be enhanced by using larger bandgap energy AlInN material (nanowires or 2D layers), and by adding surface passivation processes at the top and at the III-nitride/Si interface to reduce the top surface recombination rate and higher the shunt resistance, respectively. Moreover, the short-circuit current density could be enhanced by optimizing the contacts and reducing their resistivity to lower the series resistance.

## 4. Conclusions

In summary, we present the optimization of nanowire InN on Si(100) devices by the introduction of an a-Si buffer layer. We studied the influence of the a-Si buffer thickness (0, 4, 8, 15 and 25 nm) on the InN structural, morphological, electrical, and optical properties. Both the InN layer and Si buffer were deposited by at 550 ºC by RF and DC sputtering, respectively. The wurtzite *c*-oriented crystal structure and mosaicity of the InN remained basically unaltered by the introduction of the a-Si buffer. SEM and AFM images reveal a nanocolumnar morphology for all samples under study. The optical band gap of samples co-deposited on sapphire showed a blue-shift from ~1.8 eV to ~1.9 eV when introducing the buffer layer.

Best *n*-InN/a-Si/*p*-Si nanowire heterojunction solar cells exhibit a promising $J_{SC}$ of 17 mA/cm$^2$, $V_{OC}$ of 0.37 V and FF of 35.5% under simulated 1-sun (AM 1.5G) illumination. This result leads to an upgrading of the efficiency from 1.1% (without buffer) to 2.3% (with 15-nm buffer) due to the potential interface surface passivation of the substrate with the a-Si buffer. However, for thicker buffers the performance degrades since it prunes the carrier transportation through it. Thus, the developed structures take advantage of the combination of two effects: the enhancement of the light trapping by the nanostructured active layer and the reduction of interface parasitic effects by the a-Si buffer layer. This work suggests the first successful demonstration of InN nanowire solar cells deposited by RF sputtering and also constitutes important progress for the development of III-nitride based full-solar-spectrum photovoltaics.

## Acknowledgements

Partial financial support was provided by projects SINFOTON2-CM (P2018/NMT-4326), GRISA (CM/JIN/2021-021) and CAM-project (EPU-DPTO/2020/012). This research was also funded by MCIN/AEI/10.13039/501100011033, grant number PID2022-138434OB-C53.

The authors would also like to thank Vincent Guigoz and Aimeric Courville for their technical support for AFM and SEM measurements.